\documentclass[5p]{elsarticle}
\pdfoutput=1

\usepackage{amsfonts}
\usepackage{amsmath}
\usepackage{amssymb}
\usepackage{bm}
\usepackage{dcolumn}
\usepackage{epsfig}
\usepackage{graphicx}
\usepackage{graphics}
\usepackage[latin1]{inputenc}
\usepackage{latexsym}
\usepackage{rotating}
\usepackage{hyperref}

\journal{Physics Letters B}

\newcommand\be{\begin{equation}}
\newcommand\ba{\begin{eqnarray}}
\newcommand\ee{\end{equation}}
\newcommand\ea{\end{eqnarray}}

\newcommand{\dd}[1]{\mathrm{d}#1\,}

\begin{document}
\begin{frontmatter}
\title {A cyclic universe approach to fine tuning}
\author[dart,brown]{Stephon Alexander}
\ead{stephon.alexander@dartmouth.edu}
\author[dart]{Sam Cormack}
\ead{samuel.c.cormack.gr@dartmouth.edu}
\author[dart]{Marcelo Gleiser}
\ead{marcelo.gleiser@dartmouth.edu}

\address[dart]{Department of Physics and Astronomy, Dartmouth College
Hanover, NH 03755}
\address[brown]{Department of Physics, Brown University, Providence, RI 02906}

\date{\today}

%%%%%%%%%%%%%%%%%%%%%%%%%%%%%%%%%%%%%%%%%%%%%%%%%%%%%%%%%%%%%%%%%%%%%%%%%%%%%%%%%%%%%%%%%%%%%%
\begin{abstract}
We present a closed bouncing universe model where the value of coupling constants is set by the dynamics of a ghost-like dilatonic scalar field. We show that adding a periodic potential for the scalar field leads to a cyclic Friedmann universe where the values of the couplings vary randomly from one cycle to the next. While the shuffling of values for the couplings happens during the bounce, within each cycle their time-dependence remains safely within present observational bounds for physically-motivated values of the model parameters. Our model presents an alternative to solutions of the fine tuning problem based on string landscape scenarios.

%\noindent 
\end{abstract}
%%%%%%%%%%%%%%%%%%%%%%%%%%%%%%%%%%%%%%%%%%%%%%%%%%%%%%%%%%%%%%%%%%%%%%%%%%%%%%%%%%%%%%%%%%%%%%

%\pacs{98.80.-k,98.80.Bp,98.80.Cq}
\end{frontmatter}

%%%%%%%%%%%%%%%%%%%%%%%%%%%%%%%%%%%%%%%%%%%%%%%%%%%%%%%%%%%%%%%%%%%%%%%%%%%%%%%%%%%%%%%%%%%%%%
\paragraph{Introduction}
A fundamental problem in particle physics and cosmology concerns the specification of the constants of nature, in particular the 19 free parameters of the Standard Model. It appears that these parameters are fine-tuned to allow for the formation of complex structure and eventually life \cite{Gleiser:2012pm}. While the coupling constants of our universe are not the only ones which could lead to such structures, only some subset of all possible coupling constants could do so.  Possible solutions require new physics at high energies, as is the case with superstring theory \cite{superstrings}. For example, the Heterotic string gives rise to a four dimensional chiral gauge theory with many of the ingredients to realize the Standard Model.  However, these four dimensional compactifications present a landscape of vacua and coupling constants. The dynamics of strings in the early universe were investigated in order to build models of string cosmology \cite{Gleiser:1985,Brandenberger:1988}.  While it was the hope that string theory would univocally determine the measured couplings of the Standard Model, another approach emerged: the multiverse hypothesis \cite{multiverse,kashru}.  

Eternal inflation generically predicts that while inflation ended in our local Hubble radius, it continues in other regions, triggering the emergence of a plethora of causally-disconnected bubble universes.  If each bubble universe is endowed with different coupling constants--as generically realized in string theory--then one can use anthropic reasoning to justify the values found within our cosmic horizon, given that we are here to ask the question. This marriage between eternal inflation and the landscape of possible perturbative string compactifications provides a resolution to the  pressing question of fine tuning in modern physics. One can, however, wonder whether there are alternatives to the string landscape as a dynamical mechanism to determine the couplings of the Standard Model.  

In this work, we propose a model to explain the apparent fine-tuning of coupling constants without recourse to the multiverse.  We show that in a cyclic universe the fundamental constants can change pseudo-randomly from cycle to cycle. (We will qualify ``pseudo'' later.) Our current universe is then just the cycle which happens to contain a set of constants conducive to life. Cyclic universe models have previously been investigated as alternatives to inflation \cite{Steinhardt:2002}. The idea that different string vacua could be explored in different cycles has been suggested in the context of explaining the value of the cosmological constant \cite{Piao:2004}. A recent development in the path towards well-behaved cyclic cosmologies is the proposal of the anamorphic universe \cite{Ijjas:2015}. This approach solves the problem of anisotropic instabilities which often plague bouncing models. It also provides a mechanism for producing a nearly scale-invariant spectrum  of perturbations.

Here we will present a toy model for how a cyclic universe with pseudo-randomly changing constants might be realized.  One key ingredient is to promote all coupling constants to moduli fields, and dynamically demonstrate two features: i. During each bounce the coupling constants vary pseudo-randomly; ii. During the expansion phase in each cycle the time variation of the coupling constants remain consistent with current observational bounds. For simplicity, we will focus on the gauge sector of the Standard Model and propose how to generalize to the Yukawa sector in the conclusion.

\paragraph{The Model}
The possibility of a cyclic universe with changing constants has been investigated before \cite{Barrow:2004}. In that work, the bounce is caused by a free ghost scalar field whose kinetic energy is negative and scales as $a^{-6}$, where $a(t)$ is the FRW scale factor. The ghost dilaton field determines the value of a coupling constant, in this case the electromagnetic coupling constant. The universe is also assumed to be closed and to contain radiation. These ingredients allow for a series of closed universes separated by bounces. The value of the ghost field (and thus of the coupling) increases quickly and by the same amount during each bounce and then remains approximately constant during the following expansion/contraction cycle. The monotonically increasing coupling limits the feasibility of the model as a solution to the fine tuning problem. We note that while ghost fields remain problematic, we adopt the same phenomenological semi-classical approach as the authors in \cite{Barrow:2004} , which is to avoid its quantization.  Indeed, ghost fields have found widespread applications in field theory and cosmology, for example as candidates for phantom dark energy \cite{piazza} and k-essence inflation \cite{graham}. Additionally, in the anamorphic universe approach mentioned in the introduction, a kinetic term with the wrong sign can be rendered ghost free in the presence of a non-minimal coupling to gravity \cite{Ijjas:2015}. We are currently investigating whether our model can be embedded in the anamorphic framework and plan to report on this in future work.

Our model incorporates a potential for the ghost field in a Friedmann universe. The action is
\begin{equation}
S =\int\dd{^4x}\sqrt{-g}\left[\frac{R}{16\pi G}-\frac{1}{2}\left[\epsilon\partial_\mu \psi\partial^\mu\psi+2V(\psi)  \right] + S_{gf}
\right],
\end{equation}
with
\be
S_{gf} = -\frac{1}{4} \sum_{i} \frac{1}{(g_{YM }^{i})^2} F_{\mu\nu}^{i}F^{\mu\nu i}, 
\ee
where the coupling field for the i-th sector of the Standard Model is $g_{YM}^{i} = g_0^i e^{\psi_{i}/M_*}$, with $g_0^i$ constant, and $M_*$ some mass scale, which from here on we will take to be the Planck scale $M_p$. For clarity, we will focus on only one gauge sector; our approach is easily generalized to other sectors.
With our metric signature, $(-,+,+,+)$, $\epsilon = +1$ corresponds to a regular scalar field, while $\epsilon=-1$ corresponds to a ghost field. We take the potential to be periodic but \emph{negative},
\begin{equation}
V(\psi) = -\Lambda^4(1+\cos(\psi/f)).
\end{equation}
The negativity of the potential ensures that there is no net cosmological constant during an expansion cycle, given that the negative kinetic energy density  will drive the field to the potential maximum, where $V(\psi)=0$. The energy density and pressure of the field $\psi$ are
\begin{gather}
\rho_\psi = \frac{\epsilon}{2}\dot{\psi}^2 -\Lambda^4(1+\cos(\psi/f))\\
P_\psi = \frac{\epsilon}{2}\dot{\psi}^2 + \Lambda^4(1+\cos(\psi/f))
\end{gather}
where $f$ sets the energy scale as in axion-like models.

The equation of motion for $\psi$ in an FRW spacetime is\
\begin{equation}
\ddot{\psi} +3H\dot{\psi}-\frac{\Lambda^4}{f}\sin(\psi/f) = 0,
\label{eq:psiEOMcosmic}
\end{equation}
where $H=\dot a/a$.
We assume that other relativistic degrees of freedom are modeled by a generic radiation term, so that the Friedmann equations are
\begin{gather}
H^2 = \frac{8\pi G}{3}\left(-\frac{1}{2}\dot{\psi}^2 -\Lambda^4(1+\cos(\psi/f)) + \frac{\rho_{r0}}{a^4}  \right)-\frac{K}{a^2};\\
\label{eq:aEOMcosmic}
\frac{\ddot{a}}{a} = -\frac{8\pi G}{3}\left(-\dot{\psi}^2+\Lambda^4(1+\cos(\psi/f)) + \frac{\rho_{r0}}{a^4} \right),
\end{gather}
where $\rho_{r0}$ is the radiation energy density at $a=1$, $K=\pm 1,0$ gives the spatial curvature and we have taken $\epsilon = -1$.

The hope is that the field $\psi$ will climb onto one of the potential maxima as the universe expands so the coupling constant that it determines will not change significantly. As the universe contracts, the $\psi$ field accelerates. Its negative kinetic energy increases until it counteracts the radiation energy density and causes a bounce. At the bounce, the field is traveling quickly and can run across many maxima of the potential in both directions, resembling a sphaleron solution in electroweak baryogenesis. The precise location in the potential where it settles will set up new initial conditions for the next bounce. The field will then traverse a different number of extrema the next time there is a bounce, possibly leading to a random walk among maxima over many cycles. (Our model can evade the Tolman problem that plague cyclic universes by adding interaction terms that create entropy via the mechanism discovered in \cite{Biswas:2006bs}.)

We will work in conformal time as the bounces occur over a longer period of conformal time than cosmic time making numerical solution easier. Writing Eqs.\ \eqref{eq:psiEOMcosmic} and \eqref{eq:aEOMcosmic} in dimensionless form in terms of conformal time we have
\begin{gather}
\label{eq:psiEOMconf}
\Psi'' = -2\mathcal{H}\Psi' + \frac{a^2 \beta}{\tilde{f}}\sin(\Psi/\tilde{f});\\
\label{eq:aEOMconf}
a'' = \frac{a'^2}{a} - \frac{1}{3a} + \frac{a\Psi'^2}{3} - \frac{a^3\beta}{3}(1+\cos(\Psi/\tilde{f})),
\end{gather}
where $\Psi = \psi/M_p$, $\mathcal{H}=a'/a$, $\beta=\Lambda^4/\rho_{r0}$, $\tilde{f}=f/M_p$, and the dimensionless conformal time is $\tilde{\eta}=(\sqrt{\rho_{r0}}/M_p)\eta$, with primes denoting derivatives by $\tilde{\eta}$ and $M_p=1/\sqrt{8\pi G}$. The first Friedmann equation becomes
\begin{equation}
\label{eq:hubbleconf}
\mathcal{H}^2 = -\frac{\Psi'^2}{6} - \frac{a^2\beta}{3}\left(1+\cos(\Psi/\tilde{f})\right) +\frac{1}{3a^2} -\frac{KM_p^2}{\rho_{r0}}.
\end{equation}
When $\beta = 0$, these equations reduce to the model of Barrow et al.\ \cite{Barrow:2004} and we have exact solutions
\begin{gather}
\label{eq:psiSolnopot}
\Psi' = \frac{\sqrt{\lambda}}{a^2};\\
a^2(\eta) = \frac{1}{6}\left[1+\sqrt{1-6\lambda}\sin(\eta+\eta_0) \right],
\end{gather}
for constants $\lambda$ and $\eta_0$ depending on initial conditions. The normalization of $a$ is fixed by choosing the dimensionless curvature, $KM_p^2/\rho_{r0}= +1$. The maximum and minimum values of $a$ are
\begin{equation}
a_{\mathrm{max},\mathrm{min}} = \frac{1}{6}\left(1\pm\sqrt{1-6\lambda}\right).
\end{equation}
When $\beta = 0$ we can expand the solution about the bounce as 
\begin{equation}
\label{eq:abounce}
a(\eta) = a_\mathrm{min} \left(1+\frac{1}{2}\left(\frac{\eta}{\eta_\mathrm{bounce}}\right)^2\right),
\end{equation}
with the bounce occurring at $\eta=0$. We can plug this into Eq.\ \eqref{eq:aEOMconf} and set $\eta=0$ to get
\begin{equation}
\eta_\mathrm{bounce} = a_\mathrm{min}\sqrt{\frac{3}{1-6a_\mathrm{min}^2}} \approx \frac{a_\mathrm{min}}{a_\mathrm{max}}.
\end{equation}
This is a useful quantity since the timescale of the bounce determines how short the time steps of a numerical solver need to be in order to correctly go through the bounce. We would therefore also like to know this quantity when $\beta\neq 0$. Assuming that something like the solution in Eq.\ \eqref{eq:psiSolnopot} holds even when we include the potential, $\Psi$ moves quickly through field space at the bounce since $a$ is small. The sinusoidal term in Eq.\ \eqref{eq:psiEOMconf} therefore averages to zero during the bounce and we get back the equation of motion with no potential, whose solution is indeed given by Eq.\ \eqref{eq:psiSolnopot}. The cosine term in Eq.\ \eqref{eq:hubbleconf} also averages to zero and by setting $\mathcal{H}=0$ we get an equation for the scale factor at the bounce,
\begin{equation}
2\beta a_\mathrm{min}^6 +6a_\mathrm{min}^4-2a_\mathrm{min}^2+\lambda = 0.
\end{equation}
Note that since the solution in Eq.\ \eqref{eq:psiSolnopot} is now only valid near the bounce, the constant $\lambda$ is not as easily determined from initial conditions as it is for the case with no potential. We can, however, use this to determine the bounce time since the cosine term in equation \eqref{eq:aEOMconf} also averages to zero and we can plug in the ansatz of Eq.\ \eqref{eq:abounce} to get
\begin{equation}
\eta_\mathrm{bounce} = a_\mathrm{min}\sqrt{\frac{3}{1-6a_\mathrm{min}^2-3\beta a_\mathrm{min}^4}}, 
\label{eq:bouncetime}
\end{equation}
where we have eliminated $\lambda$. We can use this to check that for a given bounce, we are using a time step small enough to correctly capture the behavior.

Equations \eqref{eq:psiEOMconf} and \eqref{eq:aEOMconf} contain only two independent parameters, $\beta$ and $\tilde{f}$. In exploring the space of solutions we should also consider different initial conditions. Since the equations are nonlinear, the dependence of the solutions on the initial conditions will be nontrivial. There are, in principle, four initial conditions to set, $a(0)$, $a'(0)$, $\Psi(0)$ and $\Psi'(0)$. However, $a(0)$ can be fixed using the first Friedmann equation (Eq.\ \eqref{eq:hubbleconf}), and using the other initial conditions. For simplicity, we start solutions at the maximum scale factor so $a'(0)=0$ and set $\Psi(0)=\pi \tilde{f}$ so that the potential energy vanishes initially. Then the only initial condition left to vary is $\Psi'(0)$.

We solve equations \eqref{eq:psiEOMconf} and \eqref{eq:aEOMconf} numerically. We focus primarily on the behaviour of the field $\Psi$ since this determines the coupling constant in our model. We are looking for solutions with three main properties:
\begin{enumerate}
\item $\Psi$ remains approximately constant during the expansion and contraction phases.
\item $\Psi$ changes relatively quickly during the bounce phase.
\item The change in $\Psi$ can change sign from bounce to bounce in a pseudo-random way.
\end{enumerate}
These three properties allow physical constants to be approximately fixed during each cycle, but to undergo a pseudo-random walk over many cycles.

We find that in order to obtain solutions with the desired properties, we should have $\beta\sim 1$ and $\tilde{f}\ll 1$, or in terms of dimensionful quantities, $\Lambda^4\sim\rho_{r0}$ and $f\ll M_p$. For these small values of $\tilde{f}$, the field $\Psi$ crosses through many potential maxima during a bounce. Since the direction of the change in $\Psi$ during the next bounce depends sensitively on where in the potential the field ends up after the current bounce, the exact evolution becomes very sensitive to the time step used: small errors can build to the point where they change the direction of a jump in $\Psi$ which changes the subsequent evolution substantially. However, this sensitivity to the time step only affects the precise sequence of jumps and not the general behavior. As long as the time step is chosen small enough compared to the bounce times (Eq.\ \eqref{eq:bouncetime}), the numerical solution will at least be \emph{representative} of the true solution.

\begin{figure}
\centering
\includegraphics[scale=1.1]{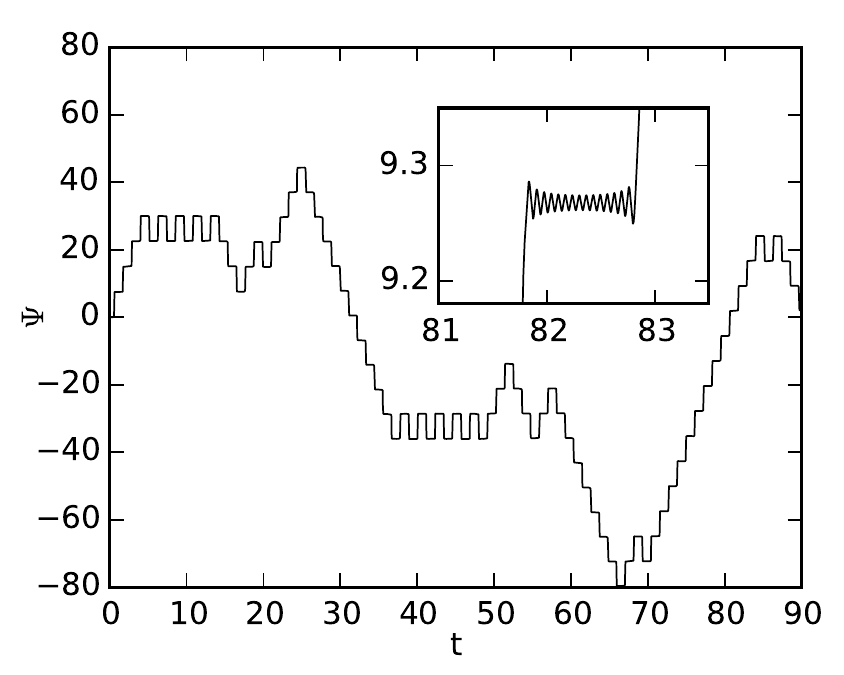}
\caption{\label{fig:psisol}The behavior of the field $\Psi$ plotted against cosmic time, $t$ (in units $M_p/\sqrt{\rho_{r0}}$). Parameters for this solution are $\beta=1$, $\tilde{f}=10^{-2}$, $\Psi'(0)=0.3$, with a time step in dimensionless conformal time $\tilde{\eta}$ of $5\times 10^{-5}$. Inset: Enlargement of the behavior of $\Psi$ between two bounces.}
\end{figure}

In Figure \ref{fig:psisol} we show an illustrative solution. The sharp changes in the field $\Psi$ correspond to bounces, while the periods where the field is comparatively constant correspond to the expansion and contraction phases. The field appears to undergo a pseudo-random walk and since it sets the coupling constant of the gauge field, $g_{YMi}$, the parameter space of coupling constants is explored over many cycles. In the inset we show the scale of oscillations in $\Psi$ during the expansion and contraction phases. We see that, as we would expect, the oscillations are of order $\tilde{f}=10^{-2}$, while the changes during the bounce have a magnitude of about 7. In fact, the magnitude of the changes during the bounce are basically independent of any parameters as long as the scale factor at the bounce is small. The scale of the oscillations away from the bounce, however, is given by the parameter $\tilde{f}$. The variation of coupling constants during the expansion and contraction phases can therefore be made arbitrarily small by choosing $\tilde{f}$ sufficiently small.

We can relate the change in $\psi$ during the expansion and contraction to the change in the coupling constant $g_{YM}$ during this time. The coupling varies as $g_{YM} = g_0 e^{\psi/M_p}=g_0 e^{\Psi}$. If $\Psi$ varies on the order of $\Delta \Psi$ and $\Delta \Psi \ll 1$ then the fractional change in the coupling constant will be of the order
\begin{equation}
\frac{\Delta g_{YM}}{g_{YM}} \sim \Delta\Psi=\frac{\Delta \psi}{M_p}.
\end{equation}
Since the parameter $f$ sets the variation of $\psi$ away from the bounce, the fractional variation of $g_{YM}$ will be of the order $f/M_p$. Observations by Webb et al.\ suggest that the fine structure constant may have varied by
\begin{equation}
\frac{\Delta \alpha}{\alpha} = -0.72 \pm 0.18 \times 10^{-5}
\end{equation}
since the early universe \cite{Webb:2001}. In our model this would require $f/M_p \sim 10^{-5}$.

During a bounce, since the value of $\psi$ changes by approximately $7M_p$, the coupling constant changes by a factor of $e^{\pm 7}\approx10^{\pm 3}$. If the bare coupling $g_0$ is of order one, then the gauge field would often become strongly coupled and could even become very strongly coupled, complicating its dynamics. However, the bare coupling may very well be many orders of magnitude smaller than one, so that even with a large change of value during a bounce, the effective theory remains safely perturbative. Given the general approach of our proposal, a viable universe--in the sense of being able to produce astrophysical structures conducive to the emergence of life--would be one where the couplings remain safely within the perturbative regime so as to emulate the Standard Model. Either way, one can assume that the majority of the contribution to the relativistic degrees of freedom remains in a thermal state, such that the energy density in radiation evolves smoothly from cycle to cycle.

\begin{figure}
\centering
\includegraphics[scale=1.1]{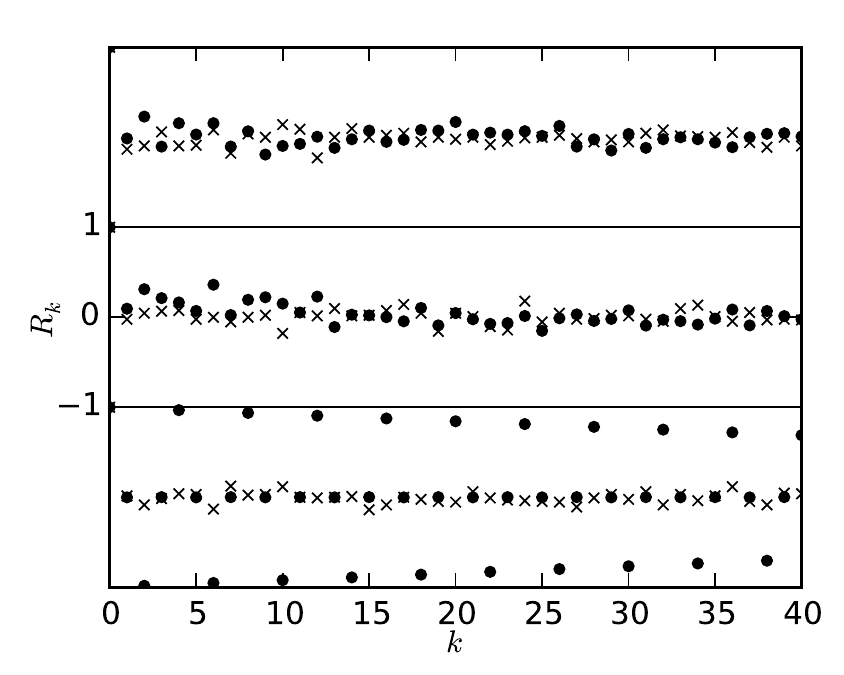}
\caption{\label{fig:autocorrelation}Autocorrelation as a function of lag $k$ for three values of $\tilde{f}$ as defined in equation \eqref{eq:autocorrelation}. From top to bottom the values of $\tilde{f}$ are $0.01$, $0.1$ and $0.3$. The circles are calculated from the model, while the crosses are for a true random walk for comparison. The $y$ axis limits are the same for the top and bottom plots as for the middle plot.}
\end{figure}

We would like to characterize the extent to which our solutions for the field $\Psi$ are well modelled by a random walk. One way to do this is to calculate the autocorrelation between differences in $\Psi$ from one cycle to the next. As a representative value of $\Psi$ from each cycle we take the value when the scale factor reaches a maximum; call this $\Psi_i$. We then take the set of differences
\begin{equation}
\Delta_i = \Psi_{i+1} - \Psi_i,
\end{equation}
and define the autocorrelation of the differences as
\begin{equation}
\label{eq:autocorrelation}
R_k = \frac{\sum_{i=1}^{N-k}\Delta_i\Delta_{i+k}}{\sum_{i=1}^N \Delta_i^2}.
\end{equation}
For a true random walk this will always be small for $k\neq 0$. We plot the autocorrelation as a function of the lag $k$ for three values of $f$ in Figure \ref{fig:autocorrelation} and compare it that for a random walk. Clearly when $f=0.3M_p$ the values are correlated (bottom plot). In fact, the solution for $\Psi$ is periodic. When $f=0.1M_p$, the autocorrelation is positive for $k$ up to around ten. For $f=0.01M_p$ (and smaller), the autocorrelation is indistinguishable from that of a random walk. This does not mean that the behavior is truly random; the dynamics are fundamentally deterministic. It does mean though, that a random walk is a good model for our solutions, and that its statistical properties will be similar. This justifies our use of the term pseudo-random walk.

It is apparent that the parameter $f$ is critical in determining the behavior of solutions. The requirement that variations in the fine structure constant are small over the lifetime of the universe tells us that $f/M_p \sim 10^{-5}$ or smaller. Note that the regime where $\tilde{f}\ll 1$ is also where the varying coupling constant is well-modeled by a random walk. While we have considered only a single gauge sector, this approach can be generalized to multiple gauge fields with independent ghost fields $\psi_i$. If these fields are not coupled (an interesting possibility), and each has a potential width parameter, $\tilde{f}_i \ll 1$, then their respective coupling constants will undergo independent pseudo-random walks. As the universe progresses through many cycles, the coupling constants will explore the parameter space.\\

\paragraph{Conclusion}
In this work we have provided an alternative cosmological model to anthropic arguments in the string landscape scenario for explaining the values of the coupling constants of the Standard Model.  Our toy model uses dilaton fields which couple to the gauge sector of the Standard Model.  We numerically demonstrated that while during the bounce the values of the coupling constants undergo a pseudo-random variation, they are stabilized during the expansion epoch of the universe.  We showed that consistency with observations naturally favors randomness. Although the mechanism stands alone as an illustration of how to implement random changes in couplings in a bounce universe, it's also motivated by string-theoretic realizations of the Standard Model where dilatons play the role of coupling constants in gauge sectors \cite{Gukov:2003cy}. A similar procedure can be implemented for Yukawa couplings, promoting them to dilaton fields with periodic potentials. As with the gauge sector, we expect them to vary pseudo-randomly during the bounce, while remaining consistent with time-dependent observational bounds during the expansion/contraction phases. We could  loosely refer to this approach as a multiverse realized in time, as one considers the variations of coupling constants over many expansion cycles. Within this framework, our cycle would be one where the couplings remain within the perturbative regime, emulating the Standard Model. 

\section*{Acknowledgement}
\noindent The authors were supported in part by a Department of Energy grant DE-SC0010386. We would like to thank Robert Caldwell for useful discussions.

\end{document}